\documentclass[twocolumn,nofootinbib]{revtex4-1}
\usepackage{latexsym,epsfig,amssymb, amsmath,nicefrac}
\usepackage{epsfig,graphicx}
\usepackage[usenames, dvipsnames]{color}
 
              \newcommand{\rf}[1]{(\ref{#1})}

\def\bfone{\relax{\rm 1\kern-.35em 1}}

\newcommand{\be}{\begin{equation}}
\newcommand{\ee}{\end{equation}}
\newcommand{\ben}{\begin{displaymath}}
\newcommand{\een}{\end{displaymath}}
\newcommand{\bea}{\begin{eqnarray}}
\newcommand{\eea}{\end{eqnarray}}

\newcommand{\bean}{\begin{eqnarray*}}
\newcommand{\eean}{\end{eqnarray*}}

\newcommand{\vp}{\varphi}

\makeatletter \@addtoreset{equation}{section} \makeatother

\begin{document}
\title{\Large{Gravitational waves and large field inflation}}

\author{Andrei Linde}

\affiliation{Department of Physics and SITP, Stanford University, \\ 
Stanford, California 94305 USA}

\begin{abstract}
According to the famous Lyth bound, one can confirm large field inflation by finding tensor modes with sufficiently large  tensor-to-scalar ratio $r$.
Here we will try to answer two related questions: Is it possible to rule out all large field inflationary models by {\it not}\, finding tensor modes with $r$ above some critical value, and what can we say about the scale of inflation by measuring $r$? However, in order to answer these questions one should distinguish between two different definitions of the large field inflation and three different definitions of the scale of inflation. We will examine these issues using the theory of cosmological $\alpha$-attractors as a convenient testing ground. 
\end{abstract}

\maketitle

\smallskip

\section{Introduction}

There is a lot of interest in the search of gravitational waves produced during inflation. Discovery of such gravitational waves would be tremendously important not only as an additional confirmation of inflationary theory, but also as a direct confirmation of predictions of quantum gravity at enormous energies. But even a non-discovery of inflationary gravitational waves would be very important since it would rule out large families of inflationary models. It already played this role by ruling out the simplest model with the quadratic potential ${m^{2}\over 2}\vp^{2}$, and by ruling out or  placing under pressure many other popular inflationary models \cite{Planck:2015xua}.

One of the tools helping to approach these issues is the famous Lyth bound \cite{Lyth:1996im}.  To derive it, one can use the slow-roll approximation and  equation for the inflaton field as a function of the number of e-foldings $N$ remaining until the end of inflation:
\be\label{slowroll}{d\vp\over dN} = {V'\over V}  \ ,
\ee
where $V'/V$ is related to the tensor to scalar ratio $r$ and the slow-roll parameter $\epsilon$,
\be
r = 16\epsilon = 8 \left({V'\over V}\right)^{2} \ .
\ee
Here and throughout the paper, we work in Planck units $M_{p} = 1$. Integrating the first equation shows that during inflation when the universe grows $e^{N}$ times the scalar field decreases by 
\be
\Delta\vp = \int_{0}^{N} dN \left({r\over 8}\right)^{1/2}  \ .
\ee
Assuming that $N > 30$ and that $r$ monotonously grows during inflation (which is often the case), one finds the bound
\be
\Delta\vp \gtrsim 10\sqrt r \ . 
\ee
One can tweak this bound a bit, but it is difficult to get around its consequences \cite{Abazajian:2016yjj}. In particular, if the cosmological observations find that $r \gtrsim 10^{{-2}}$, they will demonstrate, or at least strongly indicate, that the universe experienced large field inflation, during which the inflaton field changed by more than the Planck value, $\Delta \vp > 1$.  This result would have profound implications for the development of inflationary cosmology, and for evaluation of various theoretical ideas related to quantum gravity and string theory.

One may wonder whether it is possible to use similar considerations to rule out all large field models with $\vp > 1$ by {\it not}\, finding tensor modes with $r \gtrsim 10^{{-2}}$, or $r \gtrsim 10^{{-3}}$, or perhaps $r \gtrsim 10^{{-4}}$. The short answer to this question is ``No'', but the  long answer is more nuanced; it depends on what exactly do we mean by large field inflation. Indeed, there are at least 2 very different definitions of the large field inflation and 3 different definitions of the scale of inflation:

\vskip 5pt 

A.\,  {\it Global definition of large field inflation}.\  From the point of view of the foundations of inflationary theory, one of the main issues is whether  the 
canonically normalized 
inflaton field $\vp$ may have a super-Planckian value $\vp> 1$ {\it at any stage of the cosmological evolution},  or it is always sub-Planckian in accordance with various conjectures often debated in the literature.  

B.\,  {\it Large field inflation during the last 50-60 e-foldings}.\ From the point of view of the observational cosmology, one may want to know whether the 
canonically normalized 
field $\vp$ was large  during the last 50-60 e-foldings of inflation. 

C.\,  {\it Characteristic scale of the inflaton field}.\  One may wonder whether the characteristic scale of the inflaton field, describing a typical range $\Delta \vp$ in which the inflaton potential changes in a significant way,  can be super-Planckian. For example, the characteristic scale for the theory with a potential $V =V_{0}(1 -  e^{-\vp/M})$ as defined in   \cite{Abazajian:2016yjj} is $\Delta\vp = M$.

D.\, {\it The energy scale of inflation}. \ There is an important relation between the inflaton potential $V$ and $r$ \cite{Lyth:2009zz}
\be\label{escale1}
V^{{1/4}} \sim 1.04 \times10^{16} \ {\rm GeV} \ \left({r\over 0.01}\right)^{1/4} \ .
\ee
Some authors use this relation to argue that the discovery of the tensor modes would give us access to physics at energy scales more than $10^{11}$ times higher than those probed at the LHC collider. However, 
the comparison with the LHC collider may suggest, incorrectly, that during inflation the universe consisted of colliding particles with energies $ V^{1/4}$, comparable with the grand unification energy scale $\sim 10^{16} \ {\rm GeV}$. This is not the case in most inflationary models, where the density of elementary particles during inflation was exponentially small. The all-important exception is provided by  inflationary perturbations, which have much smaller energies, see below.

E.\, {\it The energy scale of inflationary perturbations}. \ The energy scale of these perturbations at the moment of their production is $k \sim H = \sqrt{V/3}$, in Planck mass units \cite{Linde:2005ht}. It is given by 
\be\label{hubble}
k \sim H = \sqrt{V/3} \sim 2.6\times 10^{13} \ {\rm GeV} \ \left({r\over 0.01}\right)^{1/2} \ .
\ee
It is much smaller than $V^{{1/4}}$ \rf{escale1}, but it is still a billion times higher than the energies probed at the LHC.  Yet another closely related energy scale is the Hawking temperature $T_{H}$, which is smaller than $H$ by the factor of $2\pi$.

\vskip 5pt 

Note that the definitions D and E describe {\it the energy scale} of inflation. In the rest of the paper we will concentrate on the definitions A, B and C describing {\it the scale of the inflaton field}.  

Even though the definitions A, B and C 
are related to each other, they have very different meaning. For example, the definition A is most important for investigation of general problems of inflationary cosmology, such as the problem of initial conditions for inflation. Indeed, it is much easier to solve this problem if inflation may occur at $\vp \gg 1$ \cite{Linde:1983gd,Linde:1985ub,Linde:2005ht,Carrasco:2015rva,East:2015ggf,Kleban:2016sqm,Clough:2016ymm}.  But one can have $\vp \gg 1$ at the beginning of inflation, and $\vp \ll 1$ during the last 60 e-foldings. This means that the same theory can describe large field inflation according to A, and small field inflation according to B.

Concerning the definition B, if observations tell us that $\vp > 1$ during the last 50-60 e-foldings, it will simultaneously answer  the question A, which would have fundamental significance for inflationary theory and quantum theory of gravity. Meanwhile, if we find that $\vp < 1$ during the last 50-60 e-foldings, it will rule out many inflationary models, but not the possibility that the inflaton field was large at the early stages of inflation.  

Finally, as we will see, the range of change of the field $\vp$ during the last 50-60 e-foldings of inflation can be much greater than the  characteristic scale  $M$ discussed in C. This means that the theory can be small {\it scale}\, according to C, but inflation in the same theory can be large {\it field}\, according to A and B.

One can analyze these issues using various phenomenological parametrizations of energy density or slow roll parameters in terms of the number of e-foldings $N$ \cite{Mukhanov:2013tua,Roest:2013fha,Garcia-Bellido:2014wfa,Creminelli:2014nqa}. But any parametrization is good only if the corresponding inflationary theory is well motivated \cite{Martin:2016iqo}. Therefore one may prefer to start directly with the theories, trying to find compelling models  describing  tensor modes with $r \lesssim 10^{-2}$ under the Planck constraints on $n_{s}$, and avoiding models requiring lots of fine-tuning. 

\

For a while, we did not have many ``targets of opportunity'' with $r \lesssim 10^{-2}$. The most interesting ones were three models with plateau potentials, which were quite different in nature and yet made very similar predictions:   Starobinsky model \cite{Starobinsky:1980te}, Higgs inflation  \cite{Salopek:1988qh,Bezrukov:2007ep}, and GL model of chaotic inflation in supergravity \cite{Goncharov:1983mw,Linde:2014hfa}.  During the last few years, these models were embedded into a broad class of cosmological attractors, which can describe inflation with arbitrarily small values of $r$. All of these models have similar predictions for $n_{s}$ in the small $r$ limit, providing  good match to the latest Planck data. In what follows, we will describe the simplest and yet rather general class of such models, $\alpha$-attractors \cite{Kallosh:2013hoa,Ferrara:2013rsa,Kallosh:2013yoa,Galante:2014ifa,Kallosh:2015zsa,Kallosh:2016gqp,Carrasco:2015uma,Ferrara:2016fwe}, and use these models in order to investigate relations between large field (or large scale) inflation  and tensor modes with very small $r$. 

In Section  \ref{s:attractors} we will show that the possibility of  large field inflaton at early stages of the cosmological evolution (definition A) cannot be ruled out by a non-discovery of gravitational waves with any value of $r$. Meanwhile, large field inflation during the last 50-60 e-foldings (definition B) cannot be ruled out unless one finds that $r \lesssim 2\times 10^{{-5}}$  \cite{Garcia-Bellido:2014wfa}. We will give a simple proof of this statement in Section \ref{B}. Finally, in Section \ref{C} we will show that for a broad class of $\alpha$-attractors, the characteristic scale of inflation $M$ (definition C)  \cite{Creminelli:2014nqa,Abazajian:2016yjj} is related to $r$:\, $M = N\sqrt{r/8}$, so one can measure it by finding $r$ and the number of e-foldings. This scale is super-Planckian for $r \gtrsim 2.6 \times 10^{{-3}}$, $N \sim 55$. The characteristic scale of inflation is  directly related to the geometry of the moduli space of $\alpha$-attractors \cite{Kallosh:2015zsa}.

\section{Large field inflation and $\alpha$-attractors}\label{s:attractors}

Most important  features of $\alpha$-attractors can be illustrated by a single-field  model with the Lagrangian  
 \be
 {1\over \sqrt{-g}} \mathcal{L} = { R\over 2}   -  {(\partial_{\mu} \phi)^2\over 2(1-{\phi^{2}\over 6\alpha})^{2}} - V(\phi)   \,  .
\label{cosmo}\ee
Here $\phi(x)$ is the scalar field, the inflaton.  The potentials theories of that type generalize the Starobinsky model and Higgs inflation for $\alpha = 1$ \cite{Starobinsky:1980te,Salopek:1988qh,Bezrukov:2007ep}, and GL model of chaotic inflation in supergravity \cite{Goncharov:1983mw,Linde:2014hfa} for $\alpha = 1/9$. The origin of the quadratic pole in the kinetic term can be explained in the context of hyperbolic geometry of moduli space in conformal or superconformal theory, supergravity and string theory  \cite{Kallosh:2013wya,Kallosh:2013hoa,Ferrara:2013rsa,Kallosh:2013yoa,Galante:2014ifa,Kallosh:2015zsa,Kallosh:2016gqp,Carrasco:2015uma,Ferrara:2016fwe}, or related to a non-minimal coupling of the inflaton field to gravity  \cite{Kallosh:2013wya,Kallosh:2013tua,Galante:2014ifa}.

The parameter  $\alpha$  can take any positive value. In the limit $\alpha \rightarrow \infty$ this model coincides with the standard chaotic inflation  with a canonically normalized field $\phi$ and the inflaton potential $V(\phi)$  \cite{Linde:1983gd}, with the cosmological predictions depending on the choice of $V$. However, in the most interesting case $\alpha \lesssim O(1)$, predictions of a broad class of such models converge to
 \be
  n_s  = 1-\frac{2}{N}\,, \qquad 
  r  =   \frac{12 \alpha}{N^2} \, ,  \label{aattr} 
\ee
practically independently of the choice of the potential $V$, as long as it is  non-singular and grows when the field $\phi$ approaches the boundary $\phi \to \sqrt{6\alpha}$.   These predictions    provide good fit to the Planck data  \cite{Planck:2015xua} for the often discussed range $50 < N < 60$, where $N$ is the number of e-foldings corresponding to inflationary perturbations which we can presently observe.

Since the parameter $\alpha$, and consequently $r  =  \frac{12 \alpha}{N^2}$, can take any small value without affecting $n_{s}$ much, and the predictions of these models are stable with respect to various modifications of $V(\phi)$, these models are most suitable for investigation of possible relations  between the  range of the inflaton field and the tensor to scalar ratio $r$ in the small $r$ limit. 

Many properties of $\alpha$-attractors become apparent when one makes a change of variables from the field $\phi$ to a canonically normalized inflaton field $\vp$, where 
\be\label{tanh} 
\phi = \sqrt {6 \alpha}\, \tanh{\varphi\over\sqrt {6 \alpha}} \ .
\ee
The full theory, in terms of the canonical variables, becomes
 \be
 {1\over \sqrt{-g}} \mathcal{L} = { R\over 2}   -  {(\partial_{\mu}\varphi)^{2} \over 2}  - V\big(\sqrt {6 \alpha}\, \tanh{\varphi\over\sqrt {6 \alpha}}\big)   \,  .
\label{cosmoqq}\ee
This shows that in terms of the canonically normalized field $\vp$, the potential has an infinitely long plateau, for any value of $\alpha$. 

\begin{figure}[tbp]
\centering 
\includegraphics[width=0.48\textwidth]{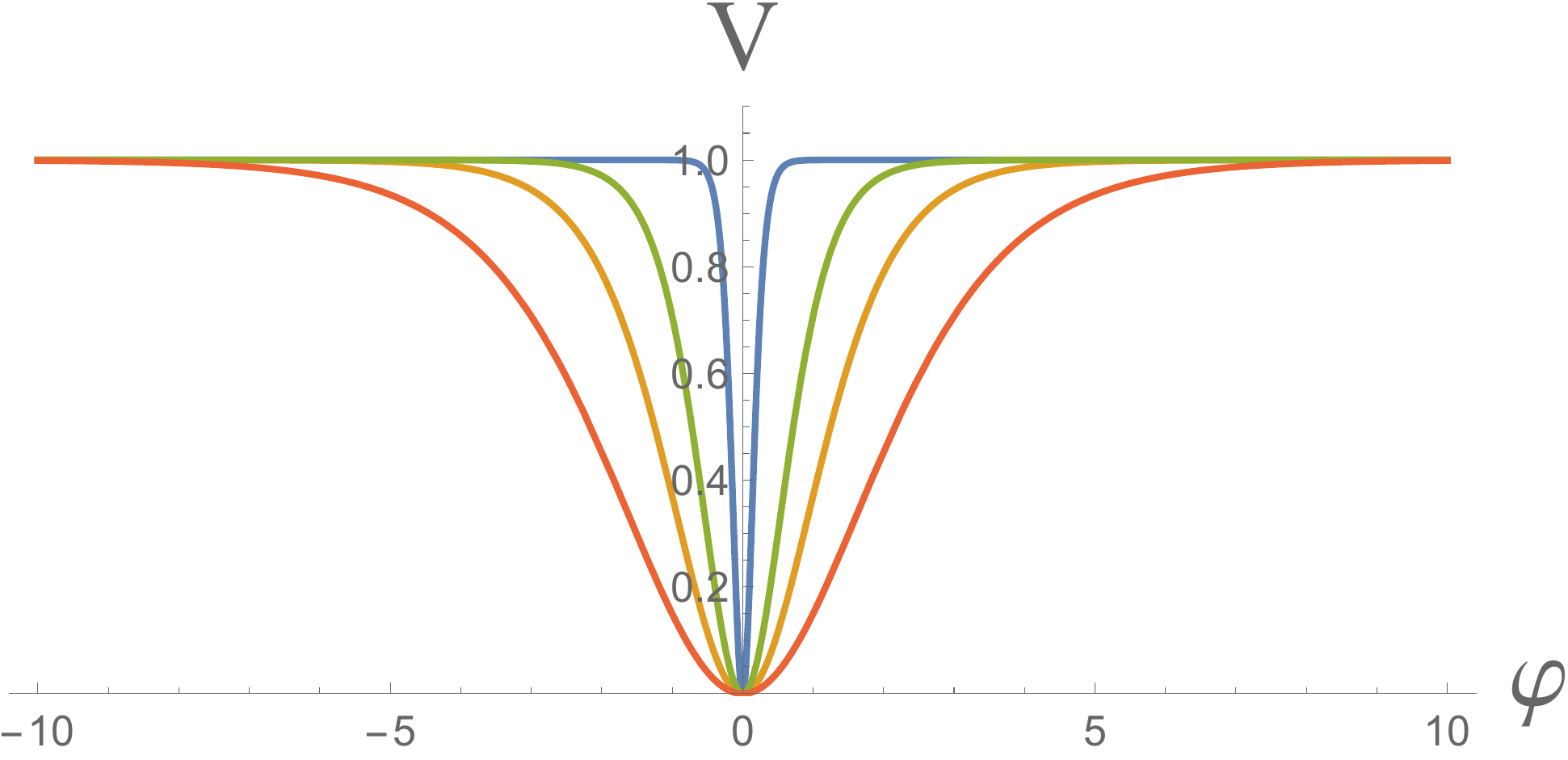}
\caption{\label{fig:1}  A family of $\alpha$-attractor potentials $V \sim  \tanh^{2}{\vp\over\sqrt {6 \alpha}}$ for various values of $\alpha$. The broad potential shown by the red line corresponds to conformal inflation \cite{Kallosh:2013hoa}, or, approximately, to  Higgs inflation \cite{Salopek:1988qh,Bezrukov:2007ep}. The next (yellow) line shows the potential for $\alpha =1/3$. It belongs to the family of potentials with $\alpha  = k/3$, $k = 1,...,7$,  suggested by geometric properties of extended supergravity/M theory/string theory \cite{Ferrara:2016fwe}.  The next (green)  line corresponds to $\alpha =1/9$, as in the GL model \cite{Goncharov:1983mw,Linde:2014hfa}. In all of these models, the value of the field $\vp$ during the last 55 e-foldings is $\vp > 1$. Finally, the internal (blue) line corresponds to $\alpha = 0.0067$, which describes the model where the value of the field $\vp$ at the beginning of the last 55 e-foldings is $\vp = 1$. In all of these models the potential is flat for $\vp \gg \sqrt{\alpha}$.}
\end{figure}

Consider, for example, potentials $V = {m^{2}}\,  \phi^{2n} $. In canonical variables, they look as follows:
\be\label{T}
 V = V_{0}   \tanh^{2n}{\vp\over\sqrt {6 \alpha}} =  V_{0}   \left(1- 4n\, e^{-\sqrt{2\over 3 \alpha}\, |\vp|}+...\right),  
  \ee
where $V_{0} = m^{2} (6\alpha)^{n}$ and ... stays for the terms  exponentially suppressed by factors higher order in $e^{-\sqrt{2\over 3 \alpha}\, |\vp|}$ \cite{Kallosh:2013hoa,Kallosh:2013yoa}. This potential has a minimum at $\vp = 0$, and an infinite flat dS plateau at $ |\vp| \gg \alpha$, see Fig. \ref{fig:1}. Flatness of this plateau is protected by the geometric properties of the theory, which result in an exponential suppression of coupling of the inflaton field to all other fields for $ |\vp| \gg \sqrt\alpha$ \cite{Kallosh:2016gqp}. 
 
Inflation can begin at any part of this infinite plateau, indefinitely far away from $\vp = 0$. It will last for exponentially long time until the field falls to the minimum of $V$ at $\vp = 0$. According to the global definition of the large field inflation (definition A),  this is an ultimate example of a  super-large-field inflation, which may occur for any $\alpha$, i.e. for any value of $r$. This leads to the following  conclusion, which we already mentioned in the previous section: 

{\it One cannot rule out the possibility of inflation at super-Plankian values of a canonically normalized inflaton field $\vp$ (i.e. of the large field inflation according to the definition A) by not finding tensor modes.}  

One can strengthen this statement even further. The broad class of cosmological attractors described above is  the simplest  class of models predicting $n_{s}$ compatible with the recent Planck data, which allows  $r$ to take arbitrary values below $10^{{-2}}$  without involving complicated and fine-tuned constructions. 

Initial conditions for inflation in such models have been studied in \cite{Carrasco:2015rva,East:2015ggf}; the conclusions of these works have been recently confirmed in  \cite{Clough:2016ymm}. One of the results of  \cite{East:2015ggf} is related to models with $\alpha,\, r \ll 1$. It was found there that the problem of initial conditions for inflation in such theories can be  solved in a rather natural way if the field $\vp$ in the early universe was very large,  $\vp > 1$.  In this sense, one may consider any  observational evidence in favor of the cosmological attractors with arbitrarily small value of $r$ as an evidence  in favor  of large field inflation.

\section{Large field inflation during the last 50-60 e-foldings}\label{B}

Now we will consider an entirely  different question: What would it take to rule out the possibility of  large field inflation with $\vp \geq 1$ {\it during the last 50-60 e-foldings of inflaton? } 
In the scenario described above, the last 50-60 e-foldings describe just a tiny part of the cosmological evolution. Therefore is not clear whether this question has any fundamental significance. But it is well formulated, and we will answer it using $\alpha$-attractors \rf{T} as an example.

Slow-roll equation  \rf{slowroll} for  the field $\vp$ at $\vp \gg \sqrt{\alpha}$ in the models \rf{T} is
\be
{d\vp\over dN} = {V'\over V} = 4n \sqrt{2\over 3 \alpha}\ e^{-\sqrt{2\over 3 \alpha}\, \vp} \ .
\ee\label{eqN}
By solving it, one finds  the following result for the value of the field $\vp_{N}$ at the beginning of the last $N$ e-foldings, in the leading approximation in $1/N$ \cite{Kallosh:2013hoa,Kallosh:2013yoa,Ueno:2016dim,Eshaghi:2016kne}:
  \be\label{field}
\vp_{N} \approx \sqrt{3 \alpha \over 2}  \log {8Nn\over 3\alpha }\ . 
\ee
One can use (\ref{field}) and find, numerically, such $\alpha$ that   $\vp_{N} = 1$ for $N = 55$, $n = 1$. The result is 
$\alpha \approx 0.0067.$
The corresponding value of $r$ is
\be
r = {12\alpha\over N^{2}}  \approx 2.6 \times 10^{{-5}} \ .
\ee
Thus for $\alpha \gtrsim 0.0067$, i.e. for $r \gtrsim 2.6 \times 10^{-5}$, the beginning of the last 55 e-folds of inflation occurs in the large field regime with  $\vp > 1$. By repeating the same analysis for other values of $n$, one finds, for example, that for   $n = 3$  the beginning of the last 55 e-folds of inflation occurs at $\vp > 1$ if  $r > 2 \times 10^{-5}$.  These  conclusions agree with the  results obtained in \cite{Garcia-Bellido:2014wfa} in a closely related context.

Our results were obtained using slow-roll approximation at $\vp \gg \sqrt\alpha$, but one can easily confirm them by solving exact equations of motion in the theory $V = V_{0}   \tanh^{2n}{\vp\over\sqrt {6 \alpha}}$, up to a small uncertainty $\Delta N \sim 1$ in the definition of what exactly is ``the end of inflation''.
Similar results for $\vp_{N}$ (up to a change $n\to n/2$ in \rf{field}) can be obtained  for the family of $\alpha$-attractors \cite{Ferrara:2013rsa,Kallosh:2013yoa} generalizing the Starobinsky model \cite{Starobinsky:1980te}, with potentials
\be
\label{E}
V=V_{0}  \left(1-e^{-{\sqrt {2\over 3 \alpha}}\vp}\right)^{2n}  . 
 \ee

Thus,  tensor modes with any value of $r$ greater than  $2 \times 10^{-5}$ are fully compatible with large field inflation during the  last 55 e-foldings. 
Moreover, following \cite{Garcia-Bellido:2014wfa}, one can formulate this statement as a strengthening of the Lyth bound: Investigation of the general class of models described above suggests that a  discovery of tensor modes with $r \gtrsim  2 \times 10^{-5}$, compatible with the Planck data for $n_{s}$, would strongly  indicate (though not prove) that the inflaton field was super-Planckian during the  last 50-60 e-foldings of inflation.

 \section{Tensor modes and characteristic scale of inflation}\label{C}
 
 All $\alpha$ attractors have plateau potentials of the type shown in equation  \rf{T}. It is convenient to represent this potential for $\vp  \gg \sqrt\alpha$ in terms of the parameter 
 \be\label{par} 
 M = \sqrt{3\alpha/ 2} \ ,
 \ee
  ignoring  terms higher order in $e^{- \vp/M}$:  
 \be\label{TM}
 V =  V_{0}   \left(1- 4n\, e^{- \vp/M}\right).  
  \ee
The parameter $M$ describes the characteristic scale  of change of the potential  $\Delta \vp = M$ \cite{Abazajian:2016yjj,Creminelli:2014nqa}.  

If one expands this potential in powers of $\vp$, one finds that
\be
 V =  V_{0} \left(1 - 4n\left(1 - {\vp\over M} +{1\over 2}\left({\vp\over M}\right)^{2} +...\right)\right) \ .
\ee
Such expressions are often interpreted as evidence suggesting that at $\phi \gtrsim M$ the theory enters strong coupling regime, with $M$ playing the role of the UV cutoff. 
Fortunately,  in the theory of $\alpha$-attractors the situation is opposite: Instead of growing up,  effective coupling constants describing strength of interaction of the field $\vp$ with itself and with all other fields  drop down as $e^{- \vp/M}$ for $\vp \gg M$. In other words, instead of the strong coupling regime with the UV cutoff $M$, one has  asymptotic freedom at $\phi \gg M$ \cite{Kallosh:2016gqp}.

It is instructive to compare  $M$ with the change of $\vp$ during the last $N$ e-foldings. Using  \rf{field}, \rf{par} one finds 
  \be\label{field2}
\vp_{N} \approx M\,  \log {4Nn\over M^{2} }\ .  
\ee
As we see from \rf{par}, for $\alpha \lesssim 1$ one has $M \lesssim 1$. Therefore for $N \gg 1$, $M \lesssim 1$ one has $\vp_{N}  \gg M$. Thus one can have  large field inflation with $\vp_{N} \gg 1$   and small scale $M \ll 1$. In particular, for $\alpha \sim 0.0067$ discussed in the previous section, one has $\vp_{55} = 1$ (large field) for $M = \sqrt{3\alpha/ 2} \approx 0.1$ (small scale).

For the broad class of $\alpha$-attractors, there is  a simple relation between $M$ and $r$, which follows from \rf{aattr}, \rf{par}:
 \be
  M = N  \sqrt {r\over 8}     \, .  \label{aattr2} 
\ee
This means that    one can determine the shape of the potential  \rf{TM} of the canonically normalized inflaton field $\vp$ by finding the  tensor-to-scalar ratio $r$.  

Equation \rf{aattr2} allows to find the constraint on $r$ corresponding to the condition $M > 1$, i.e. to $\alpha > 2/3$,  for $N = 55$: 
\be\label{critical}
r > {8 \over N^{2}} \sim 2.6 \times 10^{{-3}} \  .
\ee
For $r <  2.6 \times 10^{{-3}}$, the characteristic scale $M$ of the inflaton potential is sub-Planckian. However,  this does not mean that for $r <  2.6 \times 10^{{-3}}$ this theory describes small field inflation. Indeed, the last 55 e-foldings of inflation in the theory \rf{T} for  $M = 1$ and $n = 1$ begin at $\vp_{55} = \log{4N} \sim 5.4 \gg 1$.  Thus it is the large field inflation in accordance to the definitions A and B.

Note that the  characteristic scale  $M = \sqrt{3\alpha/2}$ describing the potential of a canonically normalized field $\vp$ coincides, up to a factor of $2$, with the maximal value $\sqrt{6\alpha}$ of the original field $\phi$ in \rf{cosmo}. Advanced versions of the theory of $\alpha$-attractors are based on supergravity describing complex scalar fields. In that case, instead of the one-dimensional interval $|\phi| < \sqrt{6\alpha}$ corresponding to the theory \rf{cosmo} one has a two-dimensional Poincar\'e disk with the  2d metric \cite{Kallosh:2015zsa}
\be
ds^2= {d r^2 + \rho^2 d\theta^{2}\over (1-{\rho^{2}\over 3\alpha})^{2}}\, , \quad \rho < \sqrt{3\alpha} \ .
\label{relat4}\ee
A Poincar\'e disk with radius $\rho = \sqrt{3\alpha}$ is a hyperbolic space with a constant negative curvature ${\cal R}_{ \mathbb{H}^2}= -{2\over 3\alpha}$ \cite{Ferrara:2013rsa,Kallosh:2013yoa}. In terms of the characteristic scale $M$, this relation looks particularly simple: 
\be
{\cal R}_{ \mathbb{H}^2}= -{M^{-2} }  \ .
\label{relat25}\ee
 This means that by finding $r = 12\alpha/N^{2}$ in this class of models one can determine the curvature of the moduli space:
\be
{\cal R}_{ \mathbb{H}^2}=   -{8\over N^{2} r }  \ .
\label{relat6}\ee 
In other words, by investigation of the universe, we can simultaneously explore geometry of the moduli space, which is responsible for the values of the inflationary parameters $n_{s}$ and $r$   \cite{Kallosh:2015zsa}.

\section{Discussion}
In this paper we explained that the possibility of the large field inflation with $\vp > 1$ at  early stages of inflation cannot be ruled out by a non-discovery of gravitational waves with any value of $r$. We also gave a simple proof of the statement that one cannot rule out large field inflation during the last 50-60 e-foldings unless one finds that $r \lesssim 2\times 10^{{-5}}$, in agreement with  \cite{Garcia-Bellido:2014wfa}. 

As for the characteristic scale of inflation $M$, we found that for a broad class of $\alpha$-attractors it is directly related to $r$ and also to the geometry of the moduli space of $\alpha$-attractors. It is super-Planckian, $M > 1$, if\, $r > {8/N^{2}} \sim 2.6 \times 10^{{-3}}$.  An important property of this scale is that the coupling constants of the canonically normalized inflaton field $\vp$ to all other fields are exponentially suppressed by the factor of $e^{- \vp/M}$ for $\vp \gg M$ \cite{Kallosh:2016gqp}. Therefore instead of the strong coupling regime with the UV cutoff $M$, one has  asymptotic freedom at $\phi \gg M$.

 It is instructive to compare the critical value $r \sim 2.6 \times 10^{{-3}}$ corresponding to $M = 1$ with the values of $r$ in the  most interesting versions of $\alpha$-attractors. For example,  GL model of chaotic inflation in supergravity \cite{Goncharov:1983mw} corresponds to $\alpha = 1/9$. For $N = 55$, which we  consider here for definiteness, it predicts $r \sim 4.4 \times 10^{{-4}}$. Conformal inflation \cite{Kallosh:2013hoa}, Starobinsky model \cite{Starobinsky:1980te} and Higgs inflation \cite{Salopek:1988qh,Bezrukov:2007ep} belong to the class with $\alpha = 1$. They have marginally super-Planckian scale of inflation $M \approx 1.2$, and predict $r \sim 4 \times 10^{{-3}}$. Considerations based on extended supergravity, M-theory and string theory suggest to pay special attention to models with $\alpha  = k/3$, $k = 1,...,7$, which would lead to $r$ in a broad range from $r \sim 1.3  \times 10^{{-3}}$  to $r \sim 0.9  \times 10^{{-2}}$ \cite{Ferrara:2016fwe}.  All of these models represent perfect targets for search of gravitational waves produced during inflation. 

Thus a detection of tensor modes with $r \gtrsim 10^{{-3}}$ would allow us to study geometry of moduli space in some of the best versions of $\alpha$-attractors, and simultaneously find out whether the scale of inflation $M$ in these models is sub-Planckian or super-Planckian. 

\

I am grateful to  J.~Carlstrom, M. Hazumi, R. Kallosh, K. Kohri  and E. Silverstein for  stimulating  discussions.   This work  is supported by the SITP,   by the NSF Grant PHY-1316699, and by the Templeton foundation grant `Inflation, the Multiverse, and Holography.'

\end{document}